\documentclass{INTERSPEECH2023}

\usepackage{amsmath,graphicx,amssymb}
\usepackage{pifont}
\usepackage{cite,threeparttable}
\usepackage{multirow,multicol,bm}
\usepackage{makecell}

\newcommand{\uvec}[1]{\boldsymbol{\hat{\textbf{#1}}}}
\newcommand{\romanOne}{\uppercase\expandafter{\romannumeral1}}
\newcommand{\romanTwo}{\uppercase\expandafter{\romannumeral2}}
\interspeechcameraready 
\title{Multi-channel multi-speaker transformer for speech recognition}
\name{Guo Yifan$^1$, Tian Yao$^1$, Suo Hongbin$^1$, Wan Yulong$^1$} 
\address{$^1$Data \& AI Engineering System, OPPO, Beijing, China}
\email{\{guoyifan, aaron1, suohongbin, wanyulong\}@oppo.com}
\begin{document}
%
\maketitle
\begin{abstract}
With the development of teleconferencing and in-vehicle voice assistants, far-field multi-speaker speech recognition has become a hot research topic.
Recently, a multi-channel transformer (MCT) has been proposed, which demonstrates the ability of the transformer to model far-field acoustic environments.
However, MCT cannot encode high-dimensional acoustic features for each speaker from mixed input audio because of the interference between speakers.
Based on these, we propose the multi-channel multi-speaker transformer (M2Former) for far-field multi-speaker ASR in this paper. Experiments on the SMS-WSJ benchmark show that the M2Former outperforms the neural beamformer, MCT, dual-path RNN with transform-average-concatenate and multi-channel deep clustering based end-to-end systems by 9.2\%, 14.3\%, 24.9\%, and 52.2\% respectively, in terms of relative word error rate reduction. 


\end{abstract}
\section{Introduction}
\vspace{-1.5mm}
\label{sec:intro}

As teleconferencing and in-vehicle voice assistants become increasingly popular, far-field multi-speaker speech recognition has become a hot research topic.
Current mainstream methods are based on the permutation invariant training (PIT) \cite{yu2017permutation}. 
Specifically, they first use far-field speech separation frontends \cite{drude2017tight, wang2018multi,chang2019mimo,luo2019conv,luo2020end,gu2020enhancing,subakan2021attention} to output single-channel speech feature for each speaker, and then decode these features with single-channel automatic speech recognition (ASR) backend with PIT. Compared to the serialized output training \cite{kanda2020serialized} based methods, PIT based methods could decode faster and are closer to practical applications. 

Recently, researchers \cite{chang2021end,chang2021multi} proposed a multi-channel transformer (MCT) for far-field speech recognition. Experiments show that MCT outperforms the commonly used systems \cite{heymann2016neural,ochiai2017multichannel,kumatani2019multi} which use enhancement models as frontends and single-channel ASR models as backends. This is mainly because the inconsistency in functional design between frontends and backends brings performance limitations to the latter systems.
Inspired by this, we try to bypass the paradigm of separation frontends recognition backends (Sep-ASR) \cite{drude2017tight, wang2018multi,chang2019mimo,luo2019conv,luo2020end,gu2020enhancing,subakan2021attention} and propose a multi-channel multi-speaker transformer (M2Former), whose encoder can encode high dimensional acoustic embeddings for each speaker from the mixture input audio directly.

However, there are certain problems in using MCT for multi-speaker scenario. 
The encoder of MCT encodes the contextual relationship under far-field environments by utilizing the intra-channel continuity information and cross-channel spatial information. 
When utilizing the cross-channel information, MCT combines channels using learnable weights which are expected to learn to model the spatial relationships implicitly.
But in the multiple speakers case, the interference between speakers and noise makes it difficult for the encoder to generate high quality acoustic embeddings for each speaker.

To alleviate such interference, two dimensional convolutional neural networks (2D-CNN) and an improved multi-channel attention mechanism called multi-channel multi-speaker attention (M2A) are used.
2D-CNN are widely used in image tasks and speech enhancement tasks \cite{sainath2015speaker,qian2016very,fu2018end,hu2020dccrn, park2020robust} as feature decoupling modules. This is due to their ability to learn diverse facets with different filter channels \cite{wei2015understanding,mahendran2015understanding,sainath2015speaker,fu2018end}.
For example, researchers \cite{sainath2015speaker} find that different filters of the CNN focus on signals coming from different directions.
Based on this, we use a 2D-CNN to decouple the multi-channel inputs first. As there are always physical or content differences between signal sources, each output channel of the CNN is expected to have a high probability of containing information corresponding to only one speaker (or noise). 
Then we propose the M2A module to encode contextual relationship. Instead of using all channels like MCT, the M2A only utilize the cross-channel information between channels with high similarity. 
Therefore, signals corresponding to different sources are encoded separately, and the interference mentioned above is avoided to a certain extent.


Furthermore, we employ a spectral clustering \cite{ng2001spectral} based approach when separating speaker-specific features from the input mixtures. Compared to the projection based method \cite{chang2019mimo,luo2019conv,luo2020end,subakan2021attention} which use linear projection layers to predict masks for speakers, there are two advantages. 1) After clustering, channels corresponding to different sound sources are assigned to different clusters. Using M2A within each cluster can better avoid the interference between sources. 2) There is no need to train additional linear layers, so it is easier to be applied to the situation of more speakers and unfixed number of speakers. Moreover, the similarity matrix required for clustering can be easily obtained from the M2A. Besides, we utilize a method \cite{guo2021far} to distinguish the noise and speech. By discarding noise cluster, the background noise is filtered out.

Based on the above approaches, we propose the M2Former for far-field multi-speaker ASR. 
Experiments on the SMS-WSJ \cite{drude2019sms} show that it outperforms the neural beamformer \cite{chang2019mimo}, MCT, dual-path RNN with transform-average-concatenate \cite{luo2020end} and multi-channel deep clustering \cite{wang2018multi} based systems by 9.2\%, 14.3\%, 24.9\%, and 52.2\% respectively, in terms of relative WER reduction.

\begin{figure*}[t]
    \centering
    \includegraphics[width=.66\textwidth]{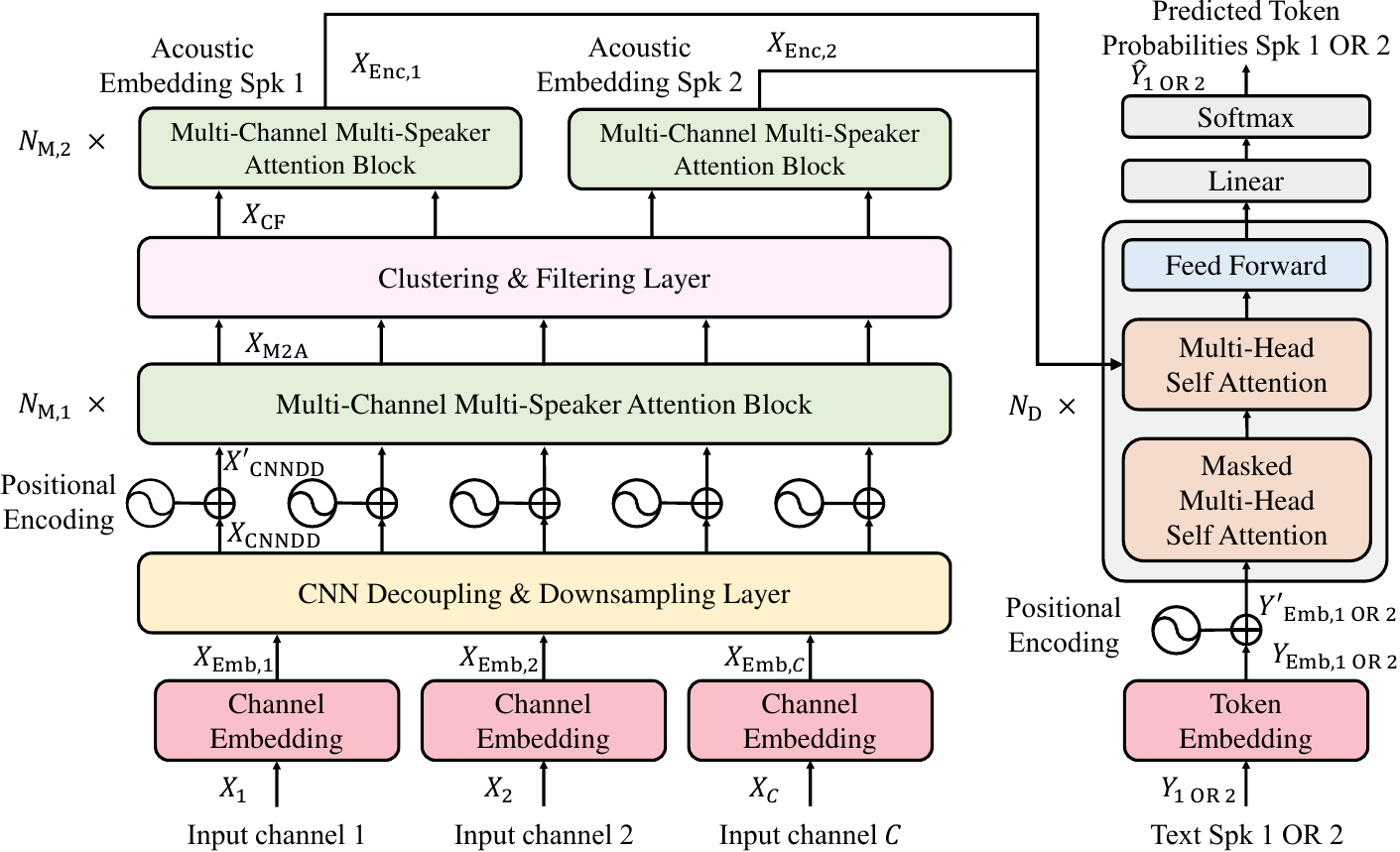} 
    \vspace{-2mm}
    \caption{Multi-channel multi-speaker transformer. $C$ is the number of input channels. $N_\text{M}$ and $ N_\text{D}$ represent the numbers of M2A blocks and decoder blocks respectively. Assume that there are two speakers in the input audio.}
    \vspace{-2.5mm}
    \label{fig:m2former}
\end{figure*}

\vspace{-1mm}
\section{Proposed Method}
\vspace{-1.5mm}
\label{sec:pagestyle}


This section describes the proposed M2Former framework, which is shown in Figure \ref{fig:m2former}. 

The structure and function of the M2Former's decoder are consistent with the decoder in the original speech transformer \cite{vaswani2017attention, dong2018speech}.
It decodes each speaker's single-channel acoustic embeddings into text separately.
Thus, we will focus on each component of the encoder, which is the core of the M2Former.

\subsection{Channel Embedding}

In order to obtain suitable input features for network learning, we add a channel embedding layer before the encoder following \cite{chang2021end}. Specifically, we use the magnitude $X^\text{mag}$ and phase features $X^\text{pha}$ of STFT as the input features $X \in \mathbb{R}^{C\times T\times 3F}$. Then we use linear projection layer to obtain the embeddings $X_\text{Emb} \in \mathbb{R}^{C\times T\times D}$ (for simplicity, we ignore the bias):
\begin{align}
    X_{\text{Emb},i} = [X_i^\text{mag}W^\text{mag},X_i^\text{pha}W^\text{pha}]\cdot W^\text{Emb}
\end{align}
Here $X_i$ denotes the $i$th channel of X, $X_{\text{Emb},i}$ denotes the $i$th channel of $X_\text{Emb}$, $W$ denotes linear projection layer, and $[\ldots, \ldots]$ denotes feature concatenation. 

\subsection{CNN Decoupling and Downsampling Layer}

Inspired by \cite{sainath2015speaker,qian2016very,fu2018end,hu2020dccrn, park2020robust}, a 2D CNN module is utilized to decouple the input multi-channel features $X_\text{Emb}$ into $X_\text{CNNDD}$ with more channels for more discriminative representations. 
Besides, motivated by the sparse distribution of the spectrum, we use the CNNs to downsample the input features on both frequency and time dimensions for computation efficiency. Thus we obtain $X_\text{CNNDD} \in \mathbb{R}^{C'\times T'\times D'}$ where $C' > C$, $T'<T$, and $D'<D$.

\subsection{Multi-Channel Multi-Speaker Attention Block}

\begin{figure*}[t]
    \centering
    \includegraphics[width=0.67\textwidth]{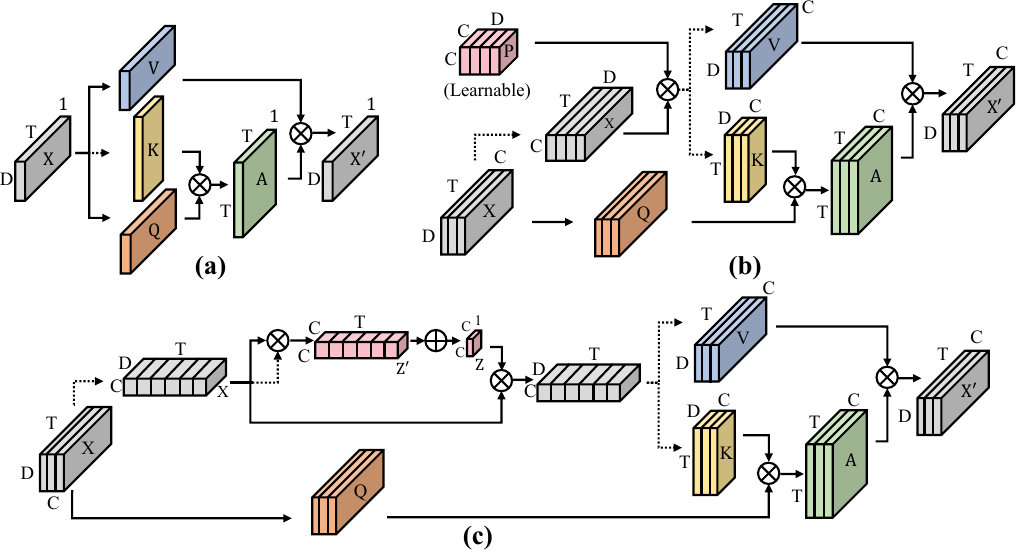} 
    \vspace{-2mm}
    \caption{Multi-channel attention blocks: \textbf{(a)}Intra-channel attention. \textbf{(b)} MCT's cross-channel attention. \textbf{(c)} M2A's cross-channel attention. $\bigotimes$, $\bigoplus$ and dashed lines denote matrix multiplication, averaging on time dimension and matrix transpose respectively.}
    \vspace{-2.5mm}
    \label{fig:multi_attention}
\end{figure*}

In order to directly encode the contextual relationship for each speaker by utilizing the intra-channel and cross-channel information, we proposed the M2A. 
Following \cite{chang2021end, chang2021multi}, the M2A consists of an intra-channel attention layer (Figure \ref{fig:multi_attention}a) and a cross-channel attention layer (Figure \ref{fig:multi_attention}c).

As described in the introduction, each output channel of the CNN is expected to have a high probability of containing information corresponding to only one speaker (or noise). So we can first utilize the self-attention in each channel in the same way as the commonly used transformer based single-channel ASR encoder does \cite{dong2018speech, chang2021end, chang2021multi} (Figure \ref{fig:multi_attention}a). It can learn the contextual continuity information within a channel. 

Then we propose a cross-channel attention layer to utilize both the continuity information between frames and the spatial information between channels. As shown in Figure \ref{fig:multi_attention}b, the MCT \cite{chang2021end,chang2021multi} combines channels using learnable weights $P$ (for simplicity, we don't introduce the specific design of the $P$). In such ways, channels belonging to different speakers would interfere with each other. This is because there are many permutations of the correspondence between channels and sound sources, which is difficult to be modeled with fixed parameters. To avoid such problem, we propose the M2A cross-channel attention (Figure \ref{fig:multi_attention}c).
Let $ X_{c}, X'_{c} \in \mathbb{R}^{T'\times D'}$ denote the $c$-th channel's input and output features of the cross-channel attention layers respectively.
Then $X'$ are computed as:
\begin{align}
\begin{split}
    &\hat{X}_{c} = \sum_{i} z_{ci} \cdot X_{i} \\
    Q_{c} &= \,\,X_{c} W^q +\mathbf{1} \cdot (\mathbf{b}^q)^\mathrm{T} \\
    K_{c} &= \,\,\hat{X}_{c} W^k +\mathbf{1} \cdot (\mathbf{b}^k)^\mathrm{T}\\
    V_{c} &= \,\,\hat{X}_{c} W^v +\mathbf{1} \cdot (\mathbf{b}^v)^\mathrm{T} \\
    X'_{c} =& \,\,
    \text{softmax}\left({Q_{c}K^\mathrm{T}_{c}}/{\sqrt{d_k}}\right)V_{c}
\end{split}
\end{align}
Where $W^*, \mathbf{b}^*$ are learnable weights and bias, $d_k$ is the scaling factor. 
We omit the multihead mechanism here for simplicity.
The $z_{ci}$ is the $c$-th row $i$-th column element of the inter-channel similarity matrix $Z \in \mathbb{R}^{C'\times C'}$, which reflects the similarity between the channel $c$ and $i$:
\begin{align}
    \begin{split}
    Z =& \,\,\,\text{softmax} \left(1/T{\sum}_{t}{X_{t}X^\mathrm{T}_{t}}/{\sqrt{d_k}}\right)
    \end{split}
\end{align}

Since the signals of different sound sources are usually uncorrelated, combining based on similarity can alleviate the interference mentioned above.
After the last M2A block in the encoder (Figure \ref{fig:m2former}), we average the channels corresponding to the same speaker to obtain a single-channel output for each speaker.

\subsection{Clustering and Filtering Layer}
\begin{figure}[t]
    \centering
    \includegraphics[width=0.43\textwidth]{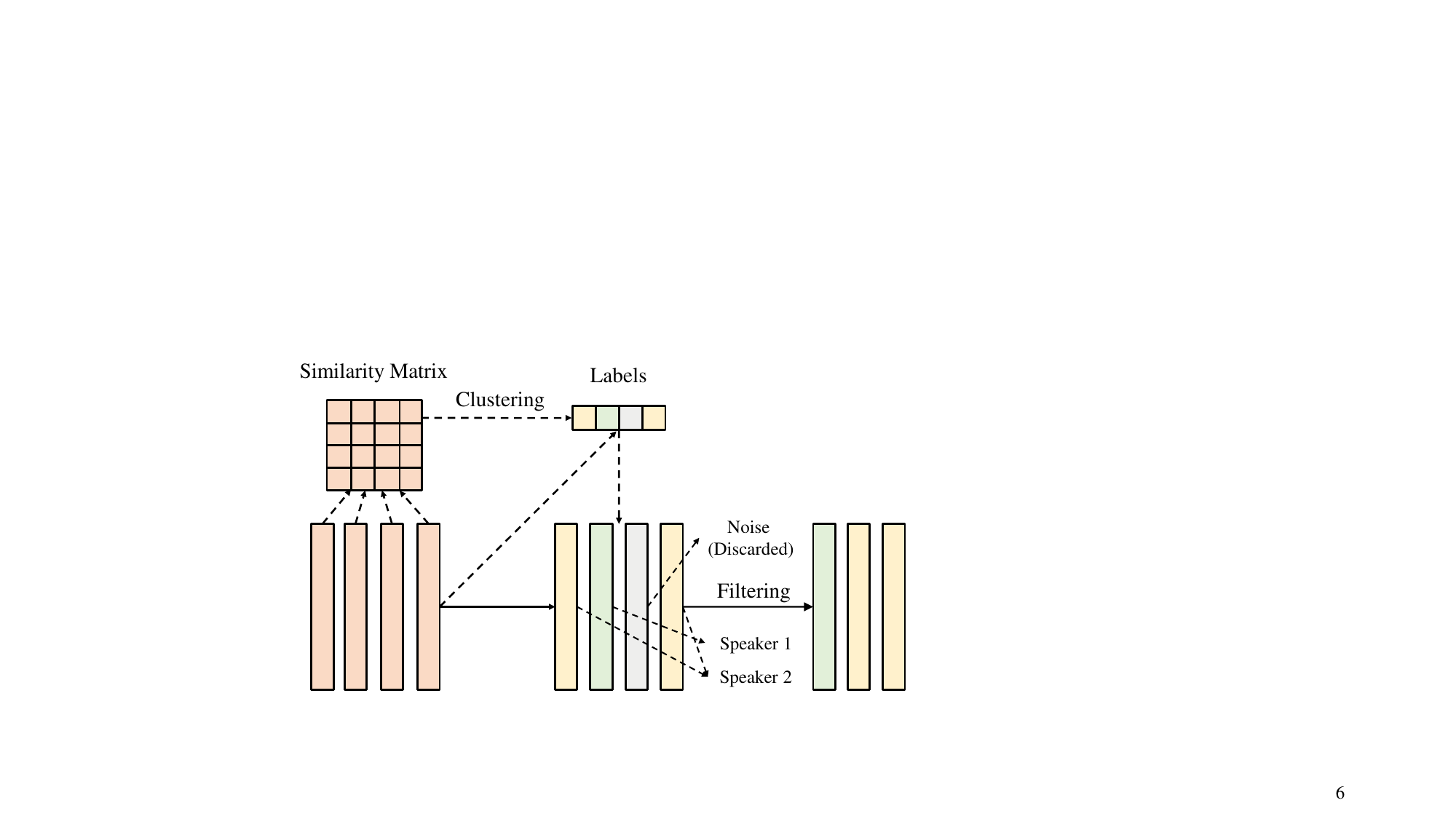} 
    \caption{Clustering and filtering layer. Gradients are propagated only in the solid line. Assume that there are 4 channels and 2 speakers, and the gray label get the lowest IFSD value.}
    \vspace{-3mm}
    \label{fig:CF}
\end{figure}

The clustering and filtering (CF) layer is shown in the figure \ref{fig:CF}. 
First, we conduct spectral clustering \cite{ng2001spectral} with the inter-channel similarity matrix $Z$ generated by the nearest M2A block to assign a label to each channel.
After that, we calculate the inter-frame similarity difference (IFSD) values \cite{guo2021far} for each output channel, 
and average the values belonging to the same label:
\begin{align}
    \text{IFSD}\left({X}\right) = \frac{1}{T} \sum_t \left[\uvec{x}_{t}^\mathrm{T}  \cdot \uvec{x}_{t+1} - \alpha \cdot \uvec{x}_{t}^\mathrm{T} \cdot \uvec{x}_{t+\tau} \right] 
    \label{eq:4}
    \vspace{-3mm}
\end{align}
Where $X \in \mathbb{R}^{D \times T}$ is the input features, $\uvec{x}_t \in \mathbb{R}^{D}$ is the normalized $t$-th frame of $X$, and $\alpha$ and $\tau$ are hyperparameters.
Finally we keep $n$ labels with highest IFSD values, and discard the others ($n$ is the number of the speakers). 
The rationale is that IFSD values represent the probabilities that 
channels are dominated by speech rather than noise. 
Thus the preserved labels correspond to the speech sources, while the  others correspond to the noise.

As shown in Figure \ref{fig:m2former}, we also try to use M2A between channels belonging to the same cluster, which get better results.

\vspace{-1mm}
\section{Experiments}
\vspace{-1mm}
\label{sec:experiment}
\subsection{Datasets}
To evaluate our proposed model, we conduct several experiments on the SMS-WSJ \cite{drude2019sms} Dataset. SMS-WSJ is a six-channel multi-speaker far-field dataset.
The SMS-WSJ has three versions of 2, 3 and 4 simultaneous speakers, and we denote them as SMS-2, SMS-3 and SMS-4. 
In this paper, we conduct experiments mainly on the SMS-2, while the other two datasets are only used for validation. The train set, valid set, and test set of SMS-2 consist of 33561, 982, and 1332 utterances respectively, for a total of 93.3 hours. SMS-3 and SMS-4 share the same original data division configuration with SMS-2. 

\vspace{-1mm}
\subsection{Configurations of Baselines and the Proposed Model}
\vspace{-0.5mm}
\subsubsection{Separation Frontends of Baselines}
To demonstrate the performance of our method, three representative multi-channel separation systems
are used as the frontends in our baselines. 
We use STFT features with 25 ms frame length and 10 ms frame shift as the raw features.
The first input channel is used as the reference channel.

\textbf{Multi-Channel Deep Clustering (MC-DPCL) \cite{wang2018multi}} : the MC-DPCL uses a 3 layer bLSTM with 300 hidden units as the embedding extractor. It takes the magnitude and phase features generated from STFT features as inputs. To make it so it can be trained end-to-end, we utilize soft k-means method as in \cite{lu2022espnet}.

\textbf{Dual-Path RNN with Transform-Average-Concatena- \\te (DPRNN-TAC) \cite{luo2020end}}: 6 DPRNN blocks are used in the system with 128 hidden units. The feature dimension and segment size of the DPRNN-TAC are 64 and 24. It takes the time domain signals as input directly.

\textbf{Neural Beamformers (NB) \cite{chang2019mimo}}: we use a 3 layer bLSTM with 300 hidden units with two linear projection layers as the mask estimators for two speakers respectively. 

\vspace{-1mm}
\subsubsection{ASR Backend of Baselines}

The ASR backend of the baselines is a unified transformer-based single-channel AED model \cite{dong2018speech} which is trained end-to-end with the frontend systems using PIT.
There are 12 blocks in the encoder, 6 blocks in the decoder, 4 heads for multi-head attention whose attention dimension is 256, and 1024 units for the position-wise feed forward. 

\vspace{-1mm}
\subsubsection{The Proposed Model}
The 2D CNN module in the CNNDD has eight CNN layers with 6, 6, 10, 10, 20, 20, 40, 40 output channels. 
The kernel sizes of the CNNs are all $3 \times 3$ with zero paddings on both dimensions. The strides of the first layer and second layer are [2,2] and [2,1], while the other layers' are all 1. $\alpha$ of IFSD is set to 5.3 according to the results on the valid set. 
To make the M2Former have the comparable size, we set the sum of $N_\text{M,1}$ and $N_\text{M,2}$ to 6, and the $N_\text{D}$ to 6, the attention dimension to 256, the heads of the multi-head attention to 4, and the feed forward units to 1024. 

\subsubsection{Loss Function of M2Former and the Baselines}
Following \cite{watanabe2017hybrid}, the loss function $\mathcal{L}$ consists of the CTC loss \cite{graves2006connectionist} of the encoder and the attention loss of the decoder. The CTC loss is used to determine the best permutation with PIT. 
\begin{align}
\begin{split}
     \mathcal{L} = {\sum}_i [\lambda  &\mathcal{L}_\text{ctc}(X_{\text{Enc}, i}, Y_{\hat{\pi}(i)})+(1-\lambda)\mathcal{L}_\text{att}(\hat{Y}_i, Y_{\hat{\pi}(i)})] \\
     \hat{\pi} &= \text{argmin}_\mathcal{{\pi}\in P}({\sum}_i\mathcal{L}_\text{ctc}(X_{\text{Enc}, i}, Y_{{\pi}(i)}))
\end{split}
\end{align}
Where $i = 1,2,...,N$ represents the speaker id, and $\mathcal{P}, Y, \hat{Y}$, $X_\text{Enc}$ represents all possible permutations, the ground truth tokens, predicted tokens and outputs of encoder respectively.

\subsection{Experimental Results}
\vspace{-0.3em}
\subsubsection{Comparison between M2Former and the Baselines}

\renewcommand\arraystretch{1}
\begin{table}[t]
  \caption{Word error rate (WER\%) of experiments on proposed model and baseline models.}
  \vspace{-0.5em}
  \centering
  \label{tab: experiments on pm and baseline}
  \begin{tabular}{lccc}
    \Xhline{2\arrayrulewidth}
    \multirow{1}{*}{SMS-2} & cv\_dev93 & test\_eval92 & \multirow{1}{*}{MEAN} \\
    \Xhline{1\arrayrulewidth}
    MC-DPCL \cite{wang2018multi} & 34.7 & 39.6 & 37.2 \\
    DPRNN-TAC \cite{luo2020end} & 22.3 & 25.1 & 23.7 \\
    NB \cite{chang2019mimo} & 18.4 & 20.7 & 19.6 \\
    \Xhline{1\arrayrulewidth}
    M2Former  & \textbf{16.6} & \textbf{18.9} & \textbf{17.8}                             \\

    \Xhline{2\arrayrulewidth}
    
  \end{tabular}
  \vspace{-.35em}
\end{table}

We conduct experiments on the SMS-2 dataset to compare the performance between the M2Former and the baselines (Table \ref{tab: experiments on pm and baseline}). Our proposed model achieves the best results on the SMS-2. Specifically, it outperforms the NB, DPRNN-TAC and MC-DPCL based end-to-end systems by 9.2\%, 24.9\%, and 52.2\% respectively, in terms of relative word error rate reduction.

\renewcommand\arraystretch{1}
\begin{table}[t]
  \caption{Ablation Experiments on the SMS-2 (WER\%).\textsl{[-]} means the module is removed during experiment).}
  \vspace{-0.5em}
  \centering
  \label{tab: ablation expe}
  \begin{tabular}{lccc}
    \Xhline{2\arrayrulewidth}
    SMS-2 &  cv\_dev93 & test\_eval92 & \multirow{1}{*}{MEAN} \\
    \Xhline{1\arrayrulewidth}
    \text{[-]} CNNDD & 35.5 & 40.1 & 37.8 \\
    \text{[-]} M2A$_1$ & 24.2 & 28.7 & 26.5 \\
    \text{[-]} M2A$_2$ & 18.3 & 21.0 & 19.7 \\
    \text{[-]} IFSD & 17.9 & 20.8 & 19.4 \\
    \Xhline{1\arrayrulewidth}
    Complete Model & \textbf{16.6} & \textbf{18.9} & \textbf{17.8} \\
    
    \Xhline{2\arrayrulewidth}
    
  \end{tabular}
   \vspace{-1.5em}
\end{table}

\subsubsection{Ablation Experiments of the Modules of the Encoder}
\label{sec: ablation}
\vspace{-0.3mm}
We conduct several ablation experiments to verify the effectiveness of each module in the encoder (Table \ref{tab: ablation expe}).
For the experiments of the CNNDD, we replace the original CNNs with two $3\times3$ 2D CNNs whose channel dimensions are 40 and strides are [2,2] and [2,1]. 
For the experiments of $\text{M2A}_1$, we just set the $N_{\text{M},2}$ to 6 and $N_{\text{M},1}$ to 0. The setup method of $\text{M2A}_2$ experiment is similar to that of the $\text{M2A}_1$, except we add a linear layer after the CF layer to smooth out the distortion introduced by the CF layer. For the IFSD experiments, we set the number of clusters to 2 and discard no channels during the CF layer.

We can find that the CNNDD contributes the most to the overall performance as it can decouple the inputs into outputs corresponding to different facets. The M2A$_1$ and the M2A$_2$ are useful. The M2A$_1$ can encode acoustic embeddings without being disturbed by other speakers. And the M2A$_2$ could eliminate the influence of the distortion caused by CF and the possible residual noise while encoding. The IFSD method also contributes to the noise cancellation.

\renewcommand\arraystretch{1}
\begin{table}[t]
  \caption{WER\% of experiments on the M2A and the MCT.}
    \vspace{-0.5em}
  \centering
  \label{tab: experiments on M2A and MCT}
  \begin{tabular}{lccc}
    \Xhline{2\arrayrulewidth}
    SMS-2 &  cv\_dev93 & test\_eval92 & \multirow{1}{*}{MEAN} \\
    \Xhline{1\arrayrulewidth}
    s-CNN + MCT & 34.8 & 38.1 & 36.5 \\
    s-CNN + STA \cite{wang2021continuous,horiguchi2022multi} & 31.4 & 35.6 & 33.6 \\ 
    s-CNN + M2A$_1$ & 40.2 & 47.7 & 44.0 \\
    \Xhline{1\arrayrulewidth}
    CNNDD + MCT & 21.2 & 24.7 & 23.0 \\
    CNNDD + STA & 19.7 & 22.8 & 21.3 \\
    CNNDD + M2A$_1$ & 18.3 & 21.0 & 19.7 \\
    
    \Xhline{2\arrayrulewidth}
    
  \end{tabular}
   \vspace{-.35em}
\end{table}

\renewcommand\arraystretch{1}
\begin{table}[t]
  \caption{Performance of M2Former with different numbers of speakers on the SMS-WSJ datasets (WER\%). The second column denotes whether the number of speakers is known in advance during inference.}
    \vspace{-0.5em}
  \centering
  \label{tab: experiments on unseen number of speakers}
  \begin{tabular}{lcccc}
    \Xhline{2\arrayrulewidth}
    


     & Known & cv\_dev93 & test\_eval92 & \multirow{1}{*}{MEAN} \\

    \Xhline{1\arrayrulewidth}

    SMS-2 & \ding{51} & 16.6 & 18.9 & 17.8\\
      & \ding{55}  & 16.9 & 19.3 & 18.1 \\
      \Xhline{0.01\arrayrulewidth}
    SMS-3 & \ding{51}  & 28.5 & 32.7 &  30.6\\
      & \ding{55}  & 30.8 & 34.5 & 32.7 \\
      \Xhline{0.01\arrayrulewidth}
    SMS-4 & \ding{51}  & 33.4 & 37.1 & 35.3 \\
      & \ding{55}  & 37.9 & 41.8 & 39.9 \\
    
    \Xhline{2\arrayrulewidth}
    
  \end{tabular}
   \vspace{-1.5em}
\end{table}

\subsubsection{Experimental comparison between the M2A and MCT}
To further explore the performance of M2A, we conduct experiments to compare the M2A with MCT and spatial-temporal attention (STA) \cite{wang2021continuous,horiguchi2022multi} (Table \ref{tab: experiments on M2A and MCT}). All 3 models use 6 attention blocks in the encoder. As the MCT can only be used with fixed number of input channels, we don't utilize attention blocks after the CF layer. We also consider the cases when replacing the CNNDD with the single CNN (s-CNN) mentioned in \ref{sec: ablation}. 

Comparing line 1 with 4 (or 2 with 5 and 3 with 6) in table \ref{tab: experiments on M2A and MCT}, the importance of the CNNDD is demonstrated again. Comparing line 1, 2 and 3, we can find that without the decoupling of the CNNDD, the M2A can be more susceptible to the interference between sources within each channel, while the MCT and STA can learn some speaker-related information implicitly. Comparing line 4, 5 and 6, it can be found that the CNNDD-M2A can better avoid interference between channels corresponding to different sound sources, and achieves a 14.3\% WER reduction relative to the CNNDD-MCT.

\vspace{-0.5mm}
\subsubsection{Performance with Different Numbers of Speakers}
\vspace{-0.3em}

We conduct a series of experiments to explore the performance of M2Former with different numbers of speakers (Table \ref{tab: experiments on unseen number of speakers}). 
The eigengap heuristic method \cite{bolla1991relations} are used to determine the number of speakers for the case where the number is unknown.

By comparing each two rows belonging to the same dataset, we can find that the number of speakers could be relatively accurately estimated from the decoupled channels.
This shows that the model has learned to decouple features to different channels according to sound sources. Besides, although the model has not been trained with more than two speakers, it works okay on the SMS-3 and SMS-4. This may show that the model could be easily applied to the situation of more speakers.

\vspace{-0.5mm}
\section{Conclusion}
\vspace{-1mm}
\label{sec:conclusion}

We proposed an end-to-end multi-channel multi-speaker transformer for speech recognition. By using CNN decoupling and downsampling layer, multi-channel multi-speaker attention block, and clustering and filtering layer, the encoder can encode speaker-wise acoustic features directly from the mixture input. The experiments shows that the proposed model outperforms the separation-recognition form baselines and the MCT.

\newpage
\bibliographystyle{IEEEtran}
\bibliography{mybib}

\end{document}